\documentclass[journal]{vgtc}                     % final (journal style)
\onlineid{0}

\vgtccategory{Research}

\vgtcpapertype{Research}

\title{Polarizing Political Polls: How Visualization Design Choices Can Shape Public Opinion and Increase Political Polarization}

\author{Eli Holder and Cindy Xiong Bearfield}
\authorfooter{ % insert punctuation at end of each item
\item
Eli Holder is with 3iap (eli@3iap.com)
\item
Cindy Xiong Bearfield is with Georgia Tech
}
\abstract{While we typically focus on data visualization as a tool for facilitating cognitive tasks (e.g.\ learning facts, making decisions), we know relatively little about their second-order impacts on our opinions, attitudes, and values. 
For example, could design or framing choices interact with viewers' social cognitive biases in ways that promote political polarization? 
When reporting on U.S. attitudes toward public policies, it is popular to highlight the gap between Democrats and Republicans (e.g.\ with blue vs red connected dot plots). But these charts may encourage social-normative conformity, influencing viewers' attitudes to match the divided opinions shown in the visualization. 
We conducted three experiments examining visualization framing in the context of social conformity and polarization. 
Crowdworkers viewed charts showing simulated polling results for public policy proposals. We varied framing (aggregating data as non-partisan “All US Adults,” or partisan “Democrat” / “Republican”) and the visualized groups' support levels. 
Participants then reported their own support for each policy. 
We found that participants' attitudes biased significantly toward the group attitudes shown in the stimuli and this can increase inter-party attitude divergence.
These results demonstrate that data visualizations can induce social conformity and accelerate political polarization. 
Choosing to visualize partisan divisions can divide us further.
}


\graphicspath{{figs/}{figures/}{pictures/}{images/}{./}} % where to search for the images

\usepackage[table,xcdraw]{xcolor}
\usepackage{tabularx}
\usepackage{tabu}                      % only used for the table example
\usepackage{booktabs}                  % only used for the table example
\usepackage{lipsum}                    % used to generate placeholder text
\usepackage{mwe}                       % used to generate placeholder figures

\usepackage{mathptmx}                  % use matching math font
\usepackage{enumitem}
\newcommand{\kw}[1]{\textbf{#1}}

\newcommand{\pheading}[1]{\vspace{4px}\noindent\textbf{#1}}
\definecolor{NavyBlue}{RGB}{32,42,6}
\definecolor{partisandotplotred}{RGB}{203,80,0}
\newcommand{\partisandotplot}[0]{\textcolor{partisandotplotred}{\kw{partisan\_dot\_plot}}}

\definecolor{partisandotrangeColor}{RGB}{203,80,0}
\newcommand{\partisandotrange}[0]{\textcolor{partisandotrangeColor}{\kw{partisan\_range}}}

\definecolor{consensusdotplotblue}{RGB}{91,141,206}
\newcommand{\consensusdotplot}[0]{\textcolor{consensusdotplotblue}{\kw{consensus\_dot\_plot}}}

\definecolor{partisantextColor}{RGB}{229,147,92}
\newcommand{\partisantext}[0]{\textcolor{partisantextColor}{\kw{partisan\_text}}}

\definecolor{controlgray}{RGB}{123,127,130}
\newcommand{\control}[0]{\textcolor{controlgray}{\kw{control}}}

\definecolor{partisanjitterplotColor}{RGB}{87,80,178}
\newcommand{\partisanjitterplot}[0]{\textcolor{partisanjitterplotColor}{\kw{partisan\_jitter\_plot}}}
\newcommand{\consensusjitterplot}[0]{\textcolor{consensusdotplotblue}{\kw{consensus\_jitter\_plot}}}

\begin{document}

% LINK TO REVISIONS: https://docs.google.com/document/d/12AUNYrtQPmPKy0sycN8sTRskhC83llmVQU_h3bvAFrQ/edit

%%%%%%%%%%%%%%%%%%%%%%%%%%%%%%%%%%%%%%%%%%%%%%%%%%%%%%%%%%%%%%%%
%%%%%%%%%%%%%%%%%%%%%% START OF THE PAPER %%%%%%%%%%%%%%%%%%%%%%
%%%%%%%%%%%%%%%%%%%%%%%%%%%%%%%%%%%%%%%%%%%%%%%%%%%%%%%%%%%%%%%%%

\firstsection{Introduction}
\maketitle

The United States is increasingly divided between Democrats and Republicans (or, more precisely, between anti-Republicans and anti-Democrats \cite{iyengar_affect_2012}). 
Americans describe their out-party peers as "close-minded" or "unintelligent" \cite{nadeem_2022} and they may be right, as increased polarization is associated with many aspects of diminished judgment \cite{prooijen_intro_2021}. 
These partisan-induced impairments impact not just citizens, political polarization is also linked with impaired governance \cite{jacobson_polarization_2016} which is a cause for concern given a widening ideology gap in Congress \cite{desilver_polarization_nodate}.

Could data visualization share some blame for this?
Previous work from social and political psychology suggests that efforts to characterize polarization can exaggerate partisanship \cite{levendusky_media_2016, rothschild_2014, madson_2018}.
With the growing popularity of data visualizations reporting polarized public opinion, differences in attitudes and norms between social groups are increasingly visible and accessible to the general public.
Perceived social norms can be influential \cite{jhangiani_conformity_2022, rothschild_2014, ledgerwood_assimilation_2007}, including through data visualization \cite{milkman_megastudies_2021, allcott_behavior_2010}. 
This is consistent with a growing body of work showing that data visualization can influence viewers' attitudes and beliefs, beyond what is strictly entailed by the data \cite{hullman_rhetoric_2011, lee-robbins_affective_2023, xiong_illusion_2020, holder_dispersion_2023, pandey2015deceptive}. 
Since polarization describes group-level attitude shifts \cite{dimaggio_polarization_1996}, considered together, these dynamics suggest a plausible path wherein visualizing political polarization may actually make it worse. 

This study explores pathways from visualization to polarization. We show that visualized attitudes can influence viewers' attitudes through social-normative conformity. 
Further, when visualizing polarized policy opinions, popular design conventions (i.e.\ red vs blue partisan dot plots) can influence readers toward more polarized attitudes. 
In some cases, this effect is stronger than reporting partisan divisions colloquially in text. 
We also explore alternative, less polarizing framings and find that, while they do not improve polarization, charts that focus on ``common ground'' at least avoid making it worse.

\pheading{Contributions:} We contribute three experiments demonstrating that data visualization can induce social conformity in readers and divergence in group-level attitudes. We tested conventional approaches to visualizing public policy opinion and showed that visualizing partisan data can increase political policy polarization.

\section{Related Work}

We discuss an interdisciplinary set of literature across social and political psychology and data visualization that inspired this work.

\subsection{Social Conformity}
We are inescapably influenced by the people around us \cite{jhangiani_conformity_2022}. If they look up, we look up \cite{milgram_1969}. If they imagine a blinking light, we imagine a blinking light \cite{sherif_psychology_1936}. If they steal ancient artifacts, we steal ancient artifacts \cite{cialdini_managing_2006}. Some argue that the very concepts that define our reality are inherently social \cite{lau_communication_2001, sloman_how_2022}. We are particularly influenceable by people who are like us, and indignant toward people who are not \cite{jhangiani_ingroup_2022}. We like people who are like us \cite{jhangiani_attraction_2022}. We show favoritism toward people inside our social identity groups (our in-groups) and prejudice toward outsiders \cite{tajfel_social_1971}. But, surprisingly, similarities do not need to be particularly meaningful to be impactful. People who share our birthdays, our hotels, letters in our first names, or our favorite books can have disproportionate influence over us \cite{burger_what_2004, goldstein_room_2008, cwir_your_2011}.

We are also susceptible to conformity when making judgments under uncertainty \cite{cosmides1996humans}. 
When interpreting data visualizations of public opinion, readers might assume the visualized opinions are (at least) as informed as their own \cite{deutsch1955study}; this implied similarity in knowledge might make these visualized opinions even more persuasive.
Prior work also demonstrates that social information about visualized data can influence readers' trust in the data, such that if data disagrees with readers' and others' prior expectations, readers see the data as less credible; in this way social information may facilitate confirmation biases \cite{kim2017explaining, kim2017data}. 
This suggests that social information, such as visualizations of public policy opinion, has the potential to influence readers' own opinions towards the visualized policies.

%%%%%%%%%%%%%%%%%%%%%%%%%%%%%%%%%%%%%%%%%%%%%%%%%%
\subsection{Polarization and Partisanship}

Polarization can be affective or ideological \cite{iyengar_affect_2012, mason_sorting_2015}. 
It can imply different distributions of people and attitudes \cite{dimaggio_polarization_1996}: 
People can sort into opposing homogeneous groups, without any individual person changing their attitudes. Or whole groups can diverge away from each other toward more extreme attitudes. Or partisan groups can converge in on themselves; by attracting ambivalent moderates away from the center, and attracting radical partisans from the wings, attitude distributions may become more bimodal, but not necessarily more extreme.
However it is defined, polarization is generally described as harmful. 
This may include stoking antipathy between fellow citizens \cite{iyengar_affect_2012, finkel_political_2020, atske_partisan_2019, iyengar_fear_2015}, creating partisan gridlock \cite{jacobson_polarization_2016}, and even afflicting polarized individuals with cognitive maladies like overconfidence, intolerance, or motivated reasoning \cite{prooijen_intro_2021}.  
Polarization can arise from individual differences \cite{federico_when_2021} or encouragement and accommodations from social actors such as policymakers or the media \cite{abramowitz_disappearing_2010, levendusky_sort_2009, klein_why_2020}.

People are sensitive to how political information is framed.
Data stories can polarize policy attitudes by pushing readers to conform with their political in-groups or diverge from political out-groups \cite{cohen_party_2003}.
For example, information with "partisan cues" (e.g.\ "Democrats support and Republicans oppose allowing prescription drug imports from Canada") can make a policy with otherwise bipartisan support substantially more popular with Democrats and less popular with Republicans \cite{clifford_increasing_2021, cohen_party_2003, malka_more_2010}. 
Priming people with partisan branding or concepts can bias beliefs and polarize political attitudes \cite{guilbeault_social_2018, ledgerwood_assimilation_2007}. 
Polling data can polarize attitudes toward candidates \cite{madson_2018}, drive support to (perceived) leading candidates \cite{rothschild_2014, boukouras_2023}, and impact voter turnout \cite{groser_2010}.
%Dichotomized election maps can increase perceived polarization and decrease voters' expected influence on elections \cite{furrer_2023}.

Social judgment theory suggests that proposed attitudes are influential in proportion to their discrepancy with prior attitudes, up to a threshold of rejection \cite{sherif_social_1961, griffin_social_2012, bochner_communicator_1966}. 
The boundaries of this rejection-threshold depend on the information's source, such that ideas that might normally be rejected become acceptable when presented by a trusted party \cite{griffin_social_2012, ledgerwood_assimilation_2007}. 
This implies that seeing the attitudes of a trusted in-group may have a conforming effect, aligned and proportionate to the distance between a viewer's starting attitude and those of the in-group. (e.g.\ Seeing that your party's attitudes are more extreme than yours might push you toward more extreme attitudes.)

This makes identifying techniques for depolarizing public policy attitudes a critical topic for investigation. For example, reducing party salience in bipartisan communication networks can reduce partisan boomerang effects in interpreting climate change data \cite{guilbeault_social_2018}. 
Highlighting incidental similarity with peers can increase receptivity to opposing viewpoints \cite{balietti_reducing_2021}. 
Priming a common national identity can decrease affective polarization \cite{levendusky_americans_2018}. 
And correcting misperceptions about out-party members can reduce negative attributions and perceived social distance \cite{lees_inaccurate_2020, ahler_parties_2018}.
As data visualizations begin to play a key role in influencing people's attitudes towards news and public policy \cite{sanchez2023effect}, we are motivated to investigate visualization techniques to communicate public opinion while minimizing the harmful effects of polarization.

%%%%%%%%%%%%%%%%%%%%%%%%%%%%%%%%%%%%%%%%%%%%%%%%%%

\subsection{Visualization and Framing Effects}

Data visualization can be influential beyond what is rationally supported by the data itself \cite{dimara2018task, luo2021attentional}. 
Visualizations can be emotional \cite{floyd_mueller_data_2021, lee-robbins_affective_2023}. The same data can be framed in ways that give rhetorical advantage to some narratives over others 
\cite{hullman_rhetoric_2011, lee-robbins_affective_2023}. 
They can combine with our prior beliefs to encourage illusory correlation \cite{xiong_seeing_2023, karduni2020bayesian}, illusory causation \cite{xiong_illusion_2020}, or motivated innumeracy \cite{baekgaard_role_2019, guilbeault_social_2018}.

Visualization design can have profound impacts on readers' interpretations and takeaways. 
The spatial arrangement of bar charts can change which data values people compare \cite{xiong2021visual, gaba2022comparison}.
Color and annotation choices can enhance objective memory of visualized content \cite{borkin2015beyond}.
Viewers' recollections of visualizations tend to align more closely with the title than the visualization itself, and title framing can lead to different perspectives \cite{kong2019trust, kong2018frames}.
Chart choices for visualizing probability distributions (e.g.\ bar histograms vs strip plots) can impact estimation accuracy for central tendency or standard deviation \cite{newburger2022fitting}. 
However, design effects related to visualizing uncertainty and variance are not limited to comprehension \cite{hofman_how_2020, kale2020visual, xiong2022investigating, correll2014error}, they can also impact trust \cite{padilla_impact_2022}, perceived effect sizes \cite{hofman_how_2020}, dichotomization fallacies \cite{wilmer_whats_2022}, and even tendencies toward harmful social stereotypes \cite{holder_dispersion_2023}.

Data visualization can play many roles in politics, such as guiding public deliberation or policy-making \cite{naerland_political_2020, franconeri_science_2021}. 
However, attitudes are not always shaped deliberately, especially in politics where group  dynamics can dominate \cite{cohen_party_2003}. 
When reading a narrative visualization \cite{segel2010narrative}, people can adjust their attitudes toward or away from the communicated message depending on how the visualized information aligns with their prior beliefs \cite{heyer2020pushing, wesslen2019investigating, xiong_seeing_2023}. 
Visualized social norms can change behaviors like gym attendance or household energy consumption \cite{milkman_megastudies_2021, allcott_behavior_2010}, influence interpretations of climate change trends \cite{guilbeault_social_2018}, and impact expectations, trust, and recall \cite{kim2017data}. 
Even seeing raw numbers of candidate favorability can polarize attitudes toward those candidates \cite{madson_2018}.
Dichotomous, adversarial comparisons between Democrats and Republicans may further exacerbate polarization by exaggerating stereotypes about political out-parties \cite{ahler_parties_2018, lees_inaccurate_2020, holder_dispersion_2023} or diminishing voters' expected influence in elections \cite{furrer_2023}.
Together, this prior work motivates our investigation into how visualizing political attitudes might channel social norms to shape public opinion.

%%%%%%%%%%%%%%%%%%%%%%%%%%%%%%%%%%%%%
\section{Study Motivation and Overview}

We conducted three experiments to understand how public opinion data visualizations can influence viewers' attitudes toward public policy, specifically investigating how visualizations can channel social-normative influences and how these combined psychological forces can contribute to political polarization. Stimuli, survey, data, and analysis scripts are available here: \href{https://osf.io/xqdw6/}{https://osf.io/xqdw6/}. 

\subsection{Hypotheses}

Building on the ideas above, we hypothesized:

\pheading{H1: Influence} Public opinion polling visualizations may have second-order impacts beyond informing viewers. They can actively influence viewers' attitudes regarding the topics being visualized.

\pheading{H2: Social Conformity} 
Polling visualizations may shift viewers' attitudes to match the visualized opinions of their relevant in-groups.
Charts showing stronger (or weaker) group support for a given policy may bias viewers toward stronger (or weaker) support.

\pheading{H3: Polarization} 
Polling visualizations' social conformity effects may increase policy polarization (i.e.\ inter-party attitude divergence).

\pheading{H4: Framing Effects} 
Visualization framing may influence polarization. 
Partisan-framed visualizations (e.g.\ blue vs red dot-plots of party-aggregate opinions) may increase polarization, while emphasizing ``common ground'' may decrease it (e.g.\ dot-plots of national-level aggregations, or jitter-plots highlighting within-party variance).

%%%%%%%%%%%%%%%%%%%%%%%%%%%%%%%%%%%%%
\subsection{Experiment Overview}

\noindent \textbf{Experiment 1}: As an exploratory study, we tested whether realistic public polling visualizations could influence attitudes on gun policies in the United States, an already highly polarized topic (\textbf{H1}).
\newline
\vspace{-2mm}

\noindent \textbf{Experiment 2}: Using a series of dynamically generated, single-issue charts, covering five different policy topics with randomly varied stimulus values representing group opinions, we investigated the influence of social conformity (\textbf{H2}), how this impacts polarization (\textbf{H3}), and partisan vs national framing effects (\textbf{H4}).
\newline
\vspace{-2mm}

\noindent \textbf{Experiment 3}: We tried, and failed, to solve political polarization with chart design (\textbf{H4}), but instead replicated our previous results.

%%%%%%%%%%%%%%%%%%%
\section{Exploratory Experiment 1: Gun-Related Policies}

In Experiment \#1, we tested whether public opinion visualizations could influence viewers' opinions (\textbf{H1}) and explored a potential pathway between visualization-triggered social conformity and political polarization (as initial investigations of \textbf{H2} and \textbf{H3}).
We used realistic stimuli to increase the external validity and real-world relevance of this initial exploration. 

\begin{figure}[!ht]
\centering
 \includegraphics[width = \columnwidth]{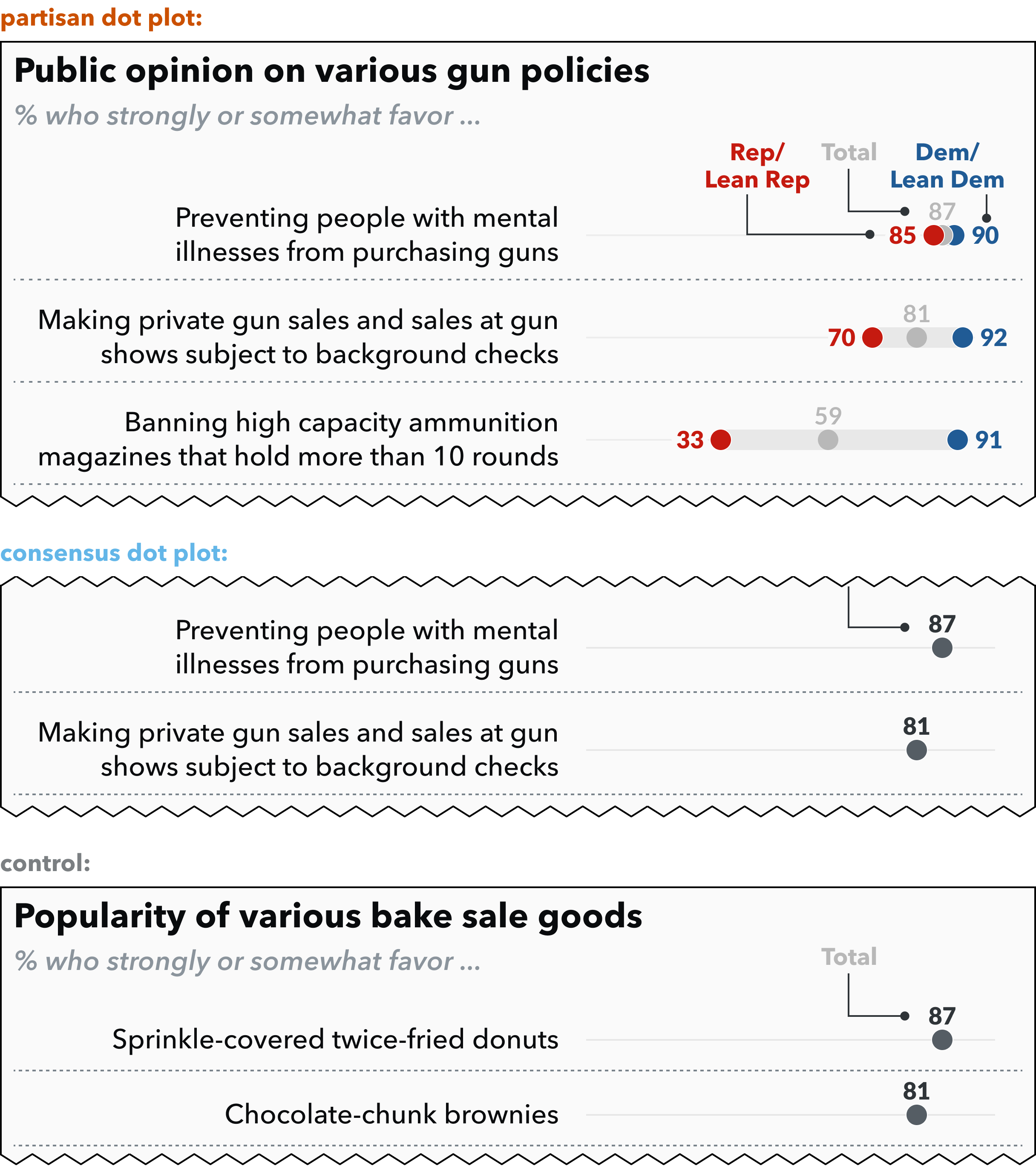}
 \caption{Experiment 1 conditions. Complete stimuli available on \href{https://osf.io/xqdw6/?view_only=8e614908f5ae42f794c650f753232c09}{OSF}.}
 \label{exp1_conditions}
\end{figure}

%%%%%%%%%%%%%%%%%%%
\subsection{Experiment 1: Design}

% \subsubsection{Experiment 1: Experiment Design}

% Experiment 1 was between-subject and post-only. 
Through a between-subject design, we asked participants to report their attitudes toward gun policies after randomly viewing one of three stimuli charts (Fig \ref{exp1_conditions}). 
We did not ask for pre-treatment attitudes because
% We did not elicit participants' attitudes before showing them the stimuli charts as a baseline in this experiment because 
%While comparing participants' attitudes before and after treatments would be more analytically straightforward (a pre-post design), 
our pilot studies and other prior work suggested this can diminish treatment effects for related (partisan cue) experiments, such that participants seek to keep their pre- and post-treatment responses consistent to avoid perceptions of partisan influence \cite{clifford_increasing_2021}.
% that for related (partisan cue) experiments, prior attitude probes create pressure to report consistent before and after responses, as participants don't wan
% that exposure to prior attitude probes might pressure participants to respond in a self-consistent fashion to avoid being perceived as non-partisan \cite{clifford_increasing_2021}, potentially diminishing effects of our manipulations due to artificial experimenter demands.
% is can lead to muted results for the particular effect studied here, as participants' desire to be seen as non-partisan creates pressure to report consistent responses \cite{clifford_increasing_2021}. 
We looked for differences between \control{} and treatment groups to estimate attitude change, using responses from politically aligned participants in the control group as a proxy for the prior attitudes of participants in the treatment groups.

\subsubsection{Experiment 1: Stimuli Design}

We tested three stimuli charts (Fig \ref{exp1_conditions}): The \consensusdotplot{} and \partisandotplot{} both show stacks of 9 gun policy proposals, with the former showing total support for the policy amongst US adults and the latter additionally showing Democrats' and Republicans' support for each policy. The \control{} shows public support for the not-yet-politicized topic of bake sale goods.

For \partisandotplot{}, the plots show plausible, but exaggerated support levels for each policy. For example, recent polling suggests 83\% of democrats support "banning assault-style weapons," but the stimulus showed 87\% \cite{pew_guns_2021}. The stimulus showed similarly exaggerated opposition, for example 37\% of Republicans support assault-weapon bans, but the stimulus showed 25\%. We chose more extreme values to increase the likely effect on people with more moderate attitudes toward each policy, based on social judgment theory's predictions that larger discrepancies between prior and proposed attitudes should lead to larger changes in attitudes \cite{bochner_communicator_1966, sherif_social_1961, griffin_social_2012}. We chose gun control because it is timely, divisive, and important, however this may have led to a more muted effect as prior work suggests that political cues are most influential for issues that are not currently politicized \cite{malka_more_2010, cohen_party_2003}. For \consensusdotplot{}, the values are based on the midpoint of the partisan values shown in \partisandotplot{}.
Each chart condition shows an "in-group" with whom participants can identify \cite{levendusky_americans_2018, ledgerwood_assimilation_2007}. For \consensusdotplot{}, we expected participants to identify with the "total" markers representing US Adults. For \partisandotplot{}, we expected participants to identify with their political party. We expect participants' reported attitudes to bias toward their respective in-groups \cite{bochner_communicator_1966, sherif_social_1961, griffin_social_2012, ledgerwood_assimilation_2007, goldstein_room_2008}. For example, for \consensusdotplot{}, attitudes should converge toward the middle of the scale regardless of participants' politics. 
For \partisandotplot{}, attitudes should bias toward participants' respective parties: For moderates, their reported attitudes should diverge from the middle, toward their parties' (relatively) more extreme attitudes. 
For more extreme partisans, their reported attitudes should converge toward the middle, toward their parties' (relatively) more moderate attitudes.

%%%%%%%%%%%%%%%%%%%
\subsection{Experiment 1: Procedure}
The experiment starts with demographic questions (age, gender, ethnicity, political affiliation) and an attention check. 
On the next page, participants are randomly assigned to see one of the three stimuli: \consensusdotplot{}, \partisandotplot{}, or \control{}.
We asked participants to answer six comprehension questions to encourage engagement with the stimuli, in addition to a simple attention check and a short free-response question where we asked for 1-2 sentences on their preferred policy or baked good, which we used as an additional attention check. % (see Participants and Exclusions). 
On the next page, participants reported their attitudes toward six specific policy statements on a continuous support scale (0 = strongly oppose, 100 = strongly favor), as shown in the \consensusdotplot{} and \partisandotplot{} stimuli in Fig. \ref{exp1_conditions}.
They also reported their attitudes toward two more general gun policy statements (e.g.\ “Greater gun restriction laws are necessary to reduce violence”).
% Attitudes were captured on a continuous support scale (0 = strongly oppose, 100 = strongly favor). 
On the last page, they reported political views as ideological alignment (0 = very liberal, 100 = very conservative), party affiliation, how strongly they identify with Democrats / Republicans (0 = not at all, 100 = very strongly), and approval of congressional Democrats / Republicans (0 = strongly disapprove, 100 = strongly approve).

%%%%%%%%%%%%%%%%%%%%%%%%%%%%%%%%%%%%%%%%%%%%%%
\subsection{Experiment 1: Participants and Exclusions}

We recruited participants on Amazon's Mechanical Turk.
We screened for people who did not participate in pilot experiments, reported locations in the United States, have an approval rating above 98\%, correctly answered two simple attention checks, provided coherent responses to the short free-response question (e.g.\ excluding responses that only say “GOOD”, or 500-word GPT outputs), and have reputable IP addresses (e.g.\ not VPNs, according to IPHub, per recommended best practices from social science research, \cite{agley2022quality, kennedy2020shape}). 
We further filtered for participants who correctly answered more than half of the chart comprehension checks and excluded independents, very liberal Republicans, or very conservative Democrats, to obtain a more comparable sample of Democrats and Republicans. Details on exclusions are available in the supplementary materials.

Mechanical Turk's worker population leans liberal and recruiting conservatives through their "Premium Qualifications" is slow and expensive. 
So we recruited toward a more balanced ratio, and ensured stimulus condition balance within party groups, but recognized that we would not achieve overall party parity. %, given time and budget constraints. 
We account for this in the study design by choosing a balanced set of policies as our stimulus and measuring political alignment on a continuous scale.
We also conducted a power analysis, based on pilot data comparing the effects of A and B ($Cohen's f$ = 0.02), suggesting that a target sample of 274 participants per condition would yield sufficient power to detect an overall difference between chart conditions, assuming an alpha level of 0.05. 

Given the above, we recruited 161 left-leaning participants and 145 right-leaning participants for \consensusdotplot{}, 183 left and 132 right participants for \partisandotplot{}, and 173 left and 133 right participants for \control{}, making up 927 participants total.

%%%%%%%%%%%%%%%%%%%%%%%%%%%%%%%%
\subsection{Experiment 1: Analysis Approach}
We predicted that visualizing public opinion would influence public opinion (\textbf{H1}). 
To test this hypothesis, we compared reported attitudes between politically aligned members of the \control{} and treatment groups.
To analyze these between-group differences, we modeled reported attitudes toward each policy (our dependent variable) as a three-way interaction between the policy's partisan alignment, the participant's partisan alignment, and the stimulus chart condition. 
Specifically, we used a linear mixed effects model, including participants' IDs as a random effect to account for individual differences. 
Each policy's partisan alignment was included as a binary variable, either left- or right- favored.
Participants' partisan alignment was included as a third-degree polynomial of their average response to the political alignment questions.  
The stimulus chart was one of: \control{}, \consensusdotplot{}, \partisandotplot{}. 
We also included an interaction between the policy's partisan alignment and factors for participants' backgrounds, to account for demographic differences in prior attitudes.
The supplement includes the detailed specification.

We modeled participants' political alignment as a  polynomial because we expected the direction of attitude change to alternate up to three times, based on the relative positions of participants' prior attitudes (their political alignment) and their in-group's reference attitude shown in the stimulus. 
A conformity effect suggests that participants will adjust their attitudes up or down to match their in-group \cite{sherif_social_1961, bochner_communicator_1966, ledgerwood_assimilation_2007}. 
So two participants, whose prior attitudes are on opposite sides of their in-group, should change their attitudes in opposite directions to conform with the group.
For example, given a plot showing US adults' support for a given policy is 57, we would expect participants' responses to pivot around that reference point. 
%\omg{(even if the plot value and the attitude scale represent different quantities)}. %XXX
Suppose a participant's prior attitude toward this policy is 19 (strong opposition), and they see that national support is 57 (moderate support). Their support should then increase, from 19 to 57 (a positive bias). 
Suppose a different participant's prior attitude toward this same policy is 83 (strong support) and they see the same plot. Their support should then decrease, from 83 to 57 (a negative bias).
In this example, both participants' attitudes converge toward the reference point (57), but they change in opposite directions given their opposite starting positions. 
Thus the direction of attitude change pivots around the reference point.
When we consider partisan charts (with both Democrat and Republican reference points) that are viewed by both liberal and conservative participants responding to their in-parties, we might expect up to three pivot points where the response curve flips directions: 1) the stimulus values shown for Democrats, 2) near x=50, which loosely delineates liberal vs conservative participants (and whether they identify with Democrat or Republican in-groups), and 3) the stimulus values shown for Republicans.

Cronbach's alpha for policy attitude questions showed consistency (liberal policies: $\alpha$=0.9, conservative: $\alpha$=0.8), with one policy more generally unpopular than the others (shorter waiting periods for gun purchases). 
Factor analysis showed that liberal and conservative policies loaded into separate factors. %, as expected. 
Political alignment measures also showed consistency ($\alpha$=0.9) and loaded into expected partisan factors. 

\begin{figure}[h!]
\centering
 \includegraphics[width = \columnwidth]{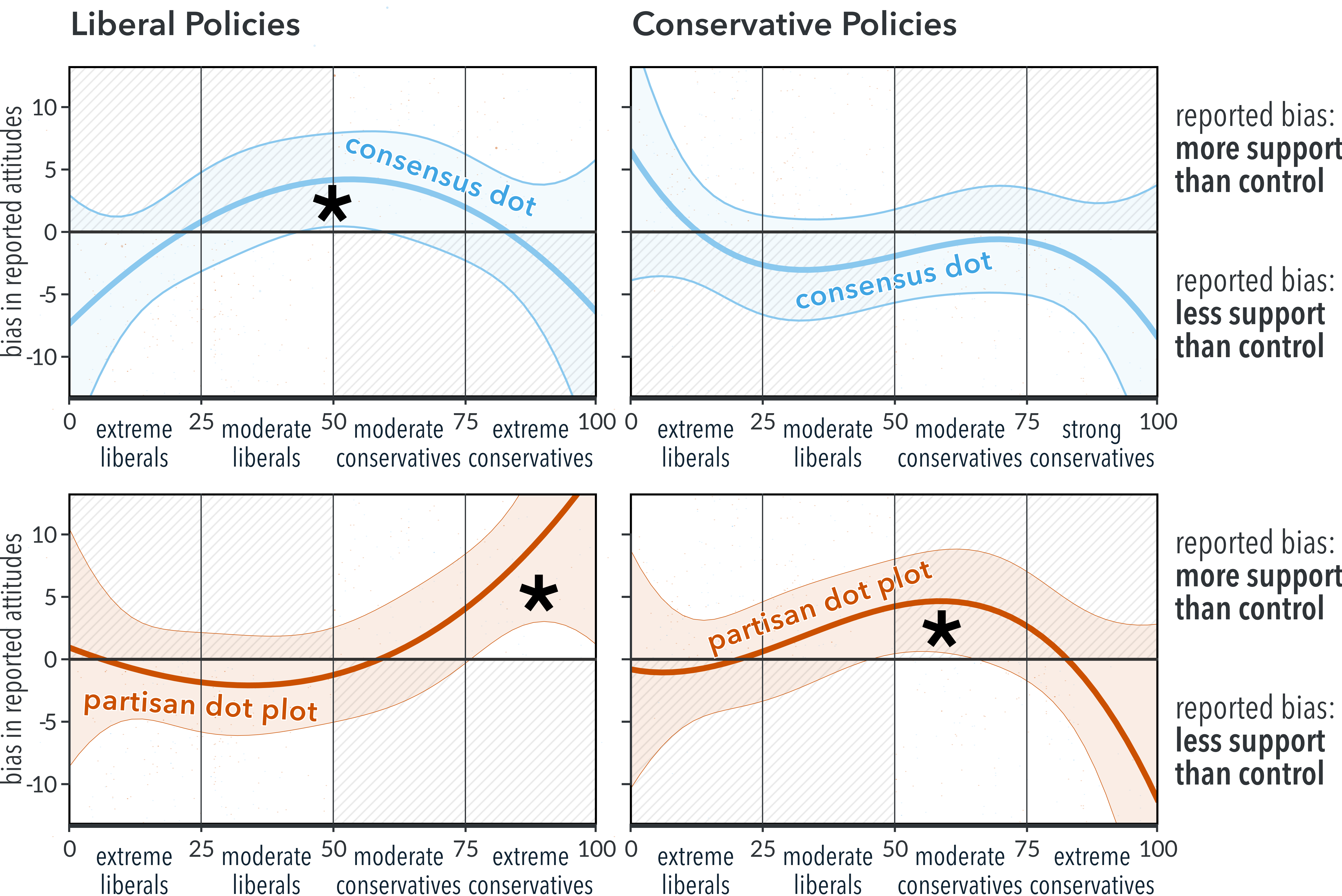}
 \caption{The partisan chart polarized moderate conservatives, increasing their support for conservative policies (bottom-right). Plots of reported attitude bias toward different sets of policies (y-axis: the difference between reported attitudes for the treatment, minus \control{}, using estimated marginal means) as a function of their political alignment (x-axis: 0 = most liberal, 100 = most conservative). Uncertainty bands indicate 95\% CIs. Positive bias indicates higher than expected support, negative indicates lower than expected. The left panels show responses to liberal policies (e.g.\ banning assault weapons), the right panels show responses to conservative policies (e.g.\ expanding concealed-carry). Cells with cross-hatching indicate polarization (i.e.\ increased support for in-party policies, or decreased support for out-party). Stars indicate significance at p<0.05.}
 \label{pol3_exp1_pol_align_to_attitude_diff}
\end{figure}

\subsection{Experiment 1: Results}

We found the expected, significant three-way interaction between chart condition, policy alignment, and participant's political alignment ($\chi^2$(6) = 29.5, p < 0.0001), supporting \textbf{H1}. 
Fig.~\ref{pol3_exp1_pol_align_to_attitude_diff} shows these relationships, as estimated marginal mean differences between the treatment conditions and \control{}. 
The supplement contains the full statistics, including model fit.
Fig.~\ref{pol3_exp1_pol_align_to_attitude_diff} shows three regions where participants' attitudes significantly differ from the \control{}.
For \consensusdotplot{} participants, the top-left panel shows that moderates increased their support for nationally-popular liberal policies (at x=[44,54], p<0.05).
For \partisandotplot{} participants,
the bottom-right panel shows that moderate-conservatives increased their support for in-party policies (at x=[46,64], p<0.05).
The bottom-left panel shows that extreme-conservatives increased their support for their out-party's policies (at x[75,100], p<0.05).

\subsection{Experiment 1: Discussion.}

In this exploratory study, we find that public opinion visualizations can influence public opinion (\textbf{H1}). % We also find early, limited support for our other hypotheses.

The observed attitude shifts are consistent with a social conformity effect (\textbf{H2}), suggesting that participants changed their attitudes toward their reference groups. 
The \consensusdotplot{} stimulus chart showed that liberal gun policies are relatively popular with US Adults. These policies became more popular with politically moderate participants who saw this chart. 
The \partisandotplot{} stimulus chart showed that conservative gun policies are supported by 44--50\% of US adults, and popular with conservatives (supported by 67--84\% of Republicans). These policies became more popular with conservative-leaning moderates who saw this chart, indicating conservatives shifting toward the conservative reference point. 
The same \partisandotplot{} stimulus showed that liberal gun policies are very popular with Democrats (supported by 87--91\%), relatively popular with US Adults, and unpopular with Republicans (but still supported by 25--33\%). These liberal policies became more popular with extreme conservatives who saw this chart. 
While this shift indicates conservatives moving in a more liberal direction, it is still consistent with social conformity when considering their prior attitudes; for the more extreme conservatives in this group, whose prior attitudes are more extreme than the Republican reference point, conforming to their in-group implies adopting a more moderate attitude.

The results show mixed support for polarization (\textbf{H3}).
Seeing the \consensusdotplot{} chart depolarized moderates' attitudes, increasing support for consensus-popular policies.
However, \partisandotplot{} had opposing effects on conservatives. It increased moderate conservatives' support for in-party policies (shifting away from consensus, consistent with polarization). It also increased extreme conservatives' support for out-party policies (shifting toward consensus, consistent with depolarization). 
The \partisandotplot{} shifts could be considered polarization based on within-party attitude consolidation \cite{dimaggio_polarization_1996}, however, they do not clearly show inter-party attitude divergence.
We explore this further in Experiment 2.

The shape of the curves in Fig.~\ref{pol3_exp1_pol_align_to_attitude_diff} may also be revealing: For moderates, the diverging biases between the treatments (top-row vs bottom-row) show different effects between the consensus and partisan framings  (\textbf{H4}).
The left-right asymmetry between liberals and conservatives may reflect differences in prior attitudes (e.g.\ liberals uniformly and strongly support gun-control policies) or asymmetry in support levels shown on the stimulus (the charts show, realistically, that liberal-favored gun policies are more generally popular).

While the results suggest a potential social-conformity effect, the story for polarization is less clear. We suspect this is related to design limitations in Experiment 1.
First, gun policy is already deeply polarized in the United States. This implies ceiling effects, where participants have little room to become \textit{more} extreme. It also implies high involvement and prior knowledge of either party's prior attitudes. These factors can make attitudes more resistant to change \cite{griffin_social_2012, cohen_party_2003}. 
Second, while \partisandotplot{} showed exaggerated attitudes for each party, they were based on realistic support levels. Realism, in this case, implies a relatively small gap between participants' prior attitudes and the group attitudes shown in the stimulus charts. This implies another possible ceiling effect, as we expect attitudes to change in proportion to the size of this gap \cite{sherif_social_1961, griffin_social_2012, bochner_communicator_1966}. We address these limitations next.

%%%%%%%%%%%%%%%%%%%%%%%%%%%%%%%%%%%%%%%%%%%
\section{Experiment 2}

We address several design limitations from Experiment 1 and further investigate the underlying processes that enable polling visualizations to lead to polarization 
We first profiled attitude change as a function of social conformity (\textbf{H2}). 
Then we investigated how these normative influences lead to the group-level shifts in attitude distributions that underlie polarization (\textbf{H3}). 
Finally, we considered differences between partisan and consensus framing effects (\textbf{H4}).

We applied several lessons-learned from Experiment 1. 
Instead of gun policies, where partisan attitudes in the U.S. are already deeply entrenched and well-known \cite{griffin_social_2012, cohen_party_2003}, we tested a mix of more obscure policies for which most participants would have little prior knowledge (or interest). 
We simplified the participant experience. Instead of reviewing a batch of 9 policies and responding on a separate page, we showed each policy individually and asked participants to respond inline, similar to prior partisan cue studies \cite{clifford_increasing_2021}. 
To clarify discrepancy-driven social conformity effects \cite{sherif_social_1961, griffin_social_2012, bochner_communicator_1966}, we switched to dynamically generated stimulus charts that randomized the values representing each reference group's support levels.

\begin{figure}[h!]
\centering
 \includegraphics[width = \columnwidth]{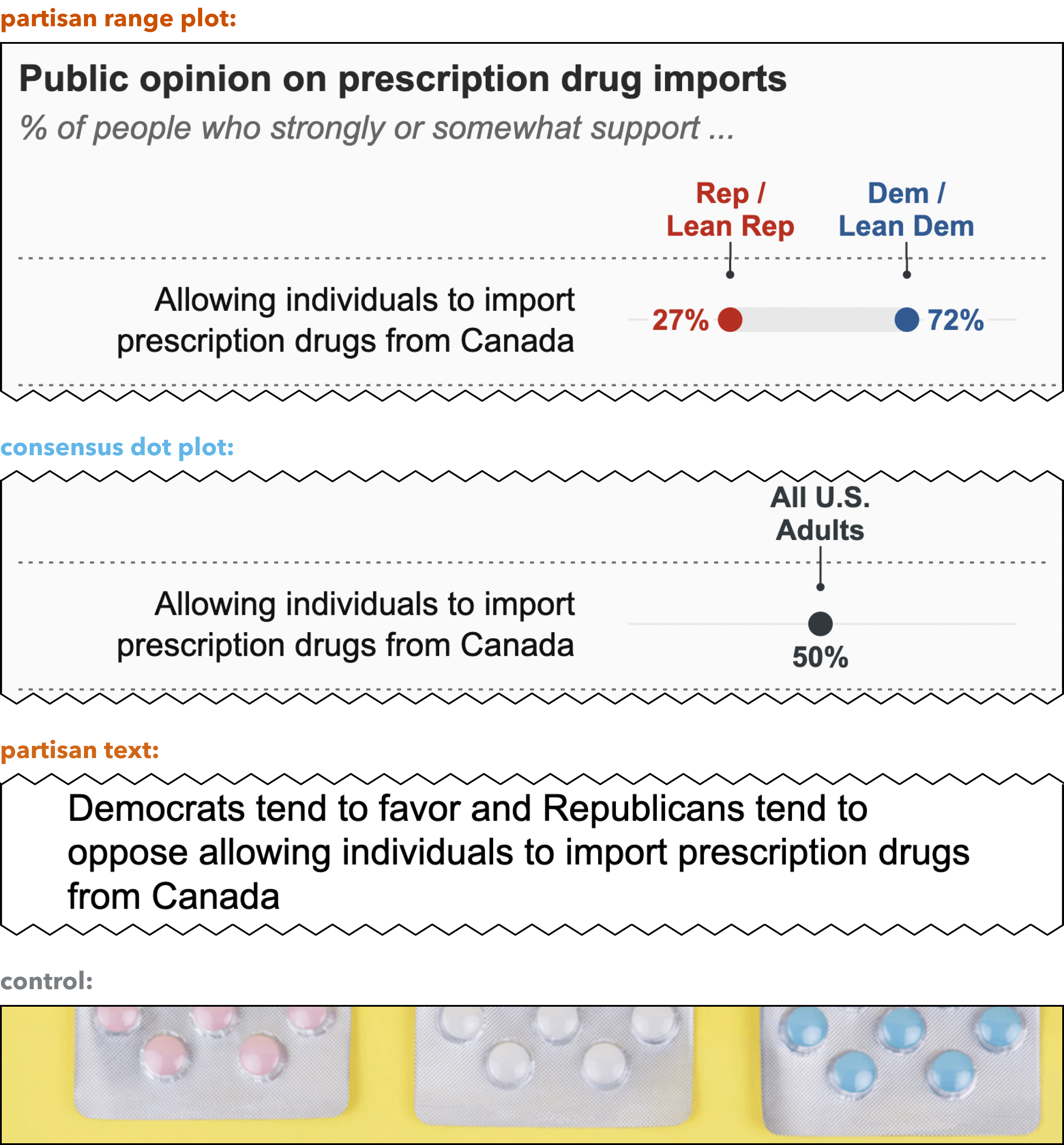}
 \caption{Experiment 2 conditions. Complete stimuli available on \href{https://osf.io/xqdw6/?view_only=8e614908f5ae42f794c650f753232c09}{OSF}.}
 \label{exp2_conditions}
\end{figure}

\subsection{Experiment 2: Design}

% \subsubsection{Experiment 2: Experiment Design}
Experiment 2 follows a similar between-subject design as Experiment 1. 
Based on our previous rationale, %attitudes are only elicited post-visualization reading. 
participants reported their attitudes toward various policies after randomly viewing one of four stimulus conditions, and we look at differences between control and treatment groups to estimate attitude change.

\subsubsection{Experiment 2: Stimuli Design} 
\label {exp2_stimuli_design_section}

We tested four stimulus chart conditions (Fig. ~\ref{exp2_conditions}).
Two are manipulation conditions: \partisandotrange{}, which showed a range between blue and red markers, representing Democrat and Republican support, and \consensusdotplot{}, which showed overall national support for ``all US adults.''
Two are control conditions, \partisantext{}, a pseudo-control which showed a verbal summary of partisan support levels (e.g.\ ``Democrats tend to favor and Republicans tend to oppose allowing individuals to import prescription drugs from Canada''), and \control{}, which showed a neutral image.
We included \partisantext{} as a control, to enable comparing our results with existing political science studies of partisan cues \cite{clifford_increasing_2021, malka_more_2010}, which use similar language to convey categorical partisan support. 
This condition also provides some limited capacity for exploring modality effects (text vs visualization), to the extent that participants' respond similarly to categorical and numeric representations of group attitudes \cite{reyna2008fuzzy}.

\pheading{Policy Topics:} We chose policy topics that were:
\begin{enumerate} [noitemsep]
\item Non-polarized (at least compared to gun control, abortion, etc), so that participants' prior attitudes were less entrenched \cite{malka_more_2010, cohen_party_2003}, 
\item Less familiar, such that other groups' support of the policy were not well known (letting us generate plausible synthetic data), 
% \item Less familiar, so that participants had little knowledge about how popular the policy is with other people (allowing us to generate plausible synthetic data), 
\item Balanced and heterogeneous in terms of plausible partisan support (e.g.\ estate taxes are unpopular with both parties, but it is plausible that conservatives dislike them more), and 
\item Across a range of baseline popularity levels, for external validity and to ensure sufficient distance between participants' prior attitudes and the simulated attitudes we showed in the charts. 
\end{enumerate}
\noindent
Based on recent prior survey work \cite{pew_driverless_2017, pew_nuclear_2022, clifford_increasing_2021}, we chose two bipartisan popular policies (prescription drug imports; self-driving cars), two bipartisanly ambivalent policies with plausible leans (estate taxes; nuclear energy), and one wildcard that could plausibly be supported by either party (genetically modified foods).

\pheading{Stimulus Values:} The charts were generated dynamically, using randomly sampled values representing (fictional, but plausible) levels of support from Democrats and Republicans, based on the following: 
\begin{enumerate} [noitemsep]
\item Values should be continuous and follow a known distribution, 
\item Most instances should show one party as clearly favoring the policy, and the other clearly opposing it (so we can observe potential polarization effects), 
\item Values should seem realistic (e.g.\ showing Republicans favoring estate taxes might raise suspicions)
% \item Topics and values should balance, such that half of the time participants see their party favoring the policy, and half their party opposes.
\item Topics and values should balance, such that participants see their party favors the policy half the time (and opposes the other half)
\end{enumerate}

\noindent
To accomplish this, we assigned four of the five topics to have fixed ``favoring'' and ``opposing'' parties (e.g.\ estate taxes and prescription drug imports typically showed support from Democrats and opposition from Republicans). GMOs were randomly assigned a favoring party.
We then sampled two values to represent either party's mean support for the policy: the ``favoring'' value was drawn from a beta distribution with mean 66, and the ``opposing'' value was from a distribution with mean 33. 
These partisan values were shown in \partisandotrange{} and \partisantext{}.
We calculated a third value representing ``All US Adults'' as the mid-point of the two partisan values. This mid-point value was shown in \consensusdotplot{}. 
We generated and tracked these stimulus values even when participants were assigned to stimuli that did not show the values directly (e.g.\ the value shown for \consensusdotplot{} was based on the generated partisan values, but does not show the partisan values; \control{} does not show any of the generated values). This lets us compare \control{} to the treatment conditions by treating the stimulus values as latent variables representing hypothetical partisan support for each policy; the values existed independently of the chart, but the treatment conditions made them visually salient.

\begin{figure*}[h!]
\centering
 \includegraphics[width = \linewidth]{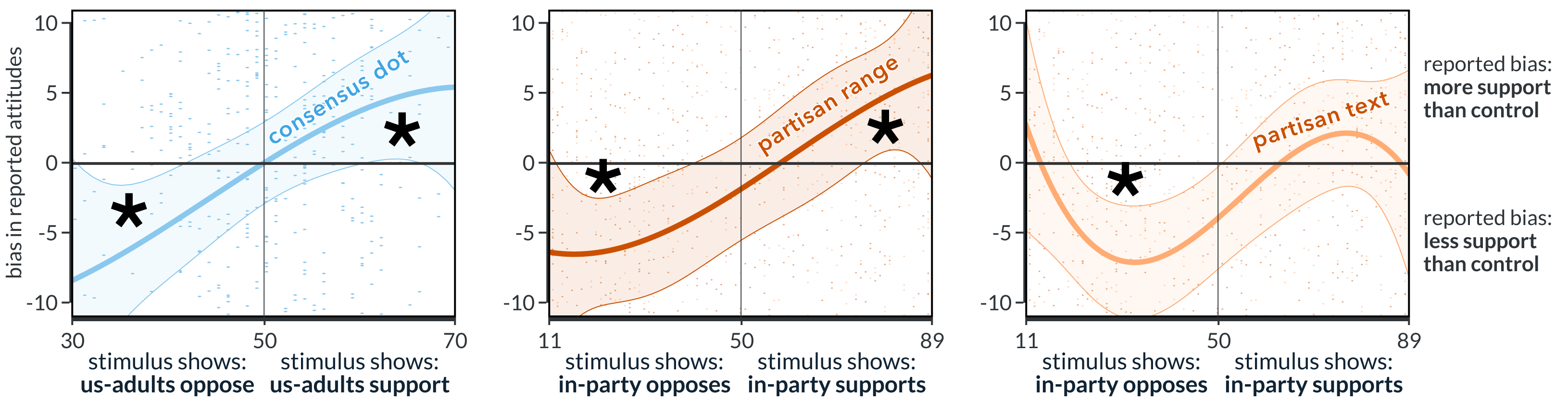}
 \caption{Experiment 2 results: Visualized attitudes influenced reported attitudes. For all three treatment conditions, participants' attitudes biased toward their visualized in-groups' attitudes. These plots show participants' bias in their reported attitudes toward various policies (y-axis: the mean difference between reported attitudes for the treatment, minus \control{}, using estimated marginal means) as a function of their in-group's visualized attitude (x-axis, 0 = in-group opposes, 100 = in-group supports). Positive bias values indicate higher than expected support, negative indicate lower than expected support. The uncertainty ranges indicate 95\% confidence intervals. Stars indicate significant differences at p<0.05.}
 \label{pol3_exp2_stim_value_vs_reported_value_mean_diff}
\end{figure*}

\subsection{Experiment 2: Procedure and Participants}

The experiment started with demographic questions, %(age, gender, ethnicity, political affiliation) 
an attention check, and how closely one follows U.S. politics. 
Participants were randomly assigned to one of the four conditions and read the five policy proposals in that condition, one on each page. 
For each policy, participants reported their attitude on a sliding scale, from 0 (strongly oppose) to 100 (strongly support). 
For the treatment conditions that contained a visualization, participants answered two comprehension questions per chart to ensure engagement. 
They lastly reported their political views as ideological alignment (0 = very liberal, 100 = very conservative), party affiliation, and how strongly they identified with Democrats and Republicans (0 = not at all, 100 = very strongly). 
We also asked about their affinity toward peer Democrats and Republicans using a feelings thermometer (0 = very cold, 100 = very warm).
The experiment ended with a debrief, explaining that the data was synthesized and should be disregarded.
We followed the same recruitment criteria from Experiment 1, resulting in 707 left-leaning and 505 right-leaning participants from MTurk, for a total of 1212 participants. Participants were balanced across chart conditions, with a similar ratio of left-to-right leaning participants within each condition.

\subsection{Experiment 2: Social Conformity}

We predicted that participants' attitudes would conform with the values shown for their in-group (\textbf{H2}), where ``in-group'' could either be ``US Adults'' or participants' reported political party. 
We again compare reported attitudes between the control and treatment groups to show the influence of each treatment. 
To analyze these between-group differences, we modeled reported attitudes toward each policy (our dependent variable) as an interaction between the stimulus chart condition and the participant's in-group's value shown in the stimulus.
Specifically, we used separate, but similarly defined linear mixed-effects models, with participants' IDs included as a random effect to account for individual differences. 
To account for participants' prior attitudes toward a given policy, we included an interaction between the policy topic and the participant's political alignment, as well as their demographic factors. 
Following our rationale from Experiment 1, we specified the in-group stimulus value and participants' political alignment as third-degree polynomials because we expected the direction of attitude change to pivot multiple times based on the relative positions of prior attitudes and the visualized reference points. 
% The supplement includes the model's complete specifications.

We modeled \partisandotrange{}, \partisantext{}, and \control{} separately from \consensusdotplot{} and \control{} because they show different reference groups, with different simulated values. 
The \partisandotrange{} and \partisantext{} charts showed simulated partisan support for each party, with values ranging between 11 and 89. The \consensusdotplot{} chart showed simulated support for ``All US Adults'', which we generated as the mid-point of the two partisan values, with values typically ranging between 30 and 70 (per \ref{exp2_stimuli_design_section}).

To evaluate conformity toward political parties, we tested \partisandotrange{} and \partisantext{} against \control{} and used the value shown for the participant's self-identified political party as the in-group stimulus value.
We found the expected significant interaction between chart condition and in-group value ($\chi^2$(6) = 33.8, p < 0.0001), indicating that participants' attitudes were influenced by the visualized attitudes of their in-party (supporting \textbf{H2}). Fig.~\ref{pol3_exp2_stim_value_vs_reported_value_mean_diff} (right two panels) shows this relationship.  We also found significant main effects for in-group value ($\chi^2$(3) = 28.7, p = p < 0.0001), confirming that participants' attitudes aligned with simulated attitudes. 
The supplement reports the full statistics and model fit.

To evaluate conformity toward a national-identity group, we tested \consensusdotplot{} against \control{}, using the value shown for ``All US Adults'' as the in-group stimulus value.
We found the expected significant interaction between chart condition and in-group value ($\chi^2$(2) = 15.0, p = 0.002), indicating that participants' attitudes were influenced by the national in-group (supporting \textbf{H2}). Fig.~\ref{pol3_exp2_stim_value_vs_reported_value_mean_diff} (first panel) shows this relationship. We also found significant main effects for in-group value ($\chi^2$(3) = 14.9, p = 0.002), confirming that participants' attitudes aligned with simulated attitudes. The supplement reports the full statistics and model fit.

As shown in Fig.~\ref{pol3_exp2_stim_value_vs_reported_value_mean_diff}, when participants in the \consensusdotplot{} or \partisandotrange{} groups saw charts showing in-group opposition to a policy (where the simulated values were less than 50), they reported less support than expected. When the charts showed in-group support, they reported more support than expected. In both cases they conformed their attitudes toward the visualized group norms. A similar pattern holds for \partisantext{}, with an apparent tapering effect.

\subsubsection{Aggregate conformity}

The previous analysis shows attitudes change as a function of in-group stimulus values, but understanding polarization means evaluating group-level shifts in attitude distributions.
For example,  Fig.~\ref{pol3_exp2_stim_value_vs_reported_value_mean_diff} shows that even when the \partisandotrange{} chart showed in-group support, it had opposing effects depending on the level of support (negative bias from x=51--57, positive bias from x=57--89). As an intermediate step toward demonstrating polarization, we examined whether these social-conformity effects aggregate to group-level shifts across the range of in-group stimulus values. While this does not necessarily imply polarization, polarization would be implausible without it.
To test this, we repeat the previous analysis with one change: instead of modeling in-group stimulus values continuously, we convert them to a categorical factor, indicating whether the in-group showed support ($>$50) or opposition ($<=$50) to each policy.
Experiment 2 was designed to support these categorical comparisons.
Even though the in-group stimulus values were randomly generated, they were sampled from distributions centered around 33 (oppose) or 66 (support) to balance the overall instances where participants saw their in-groups supporting or opposing each policy. 
This lets us examine aggregate conformity across a range of visualized attitudes.

\begin{figure}[h!]
\centering
 \includegraphics[width = \columnwidth]{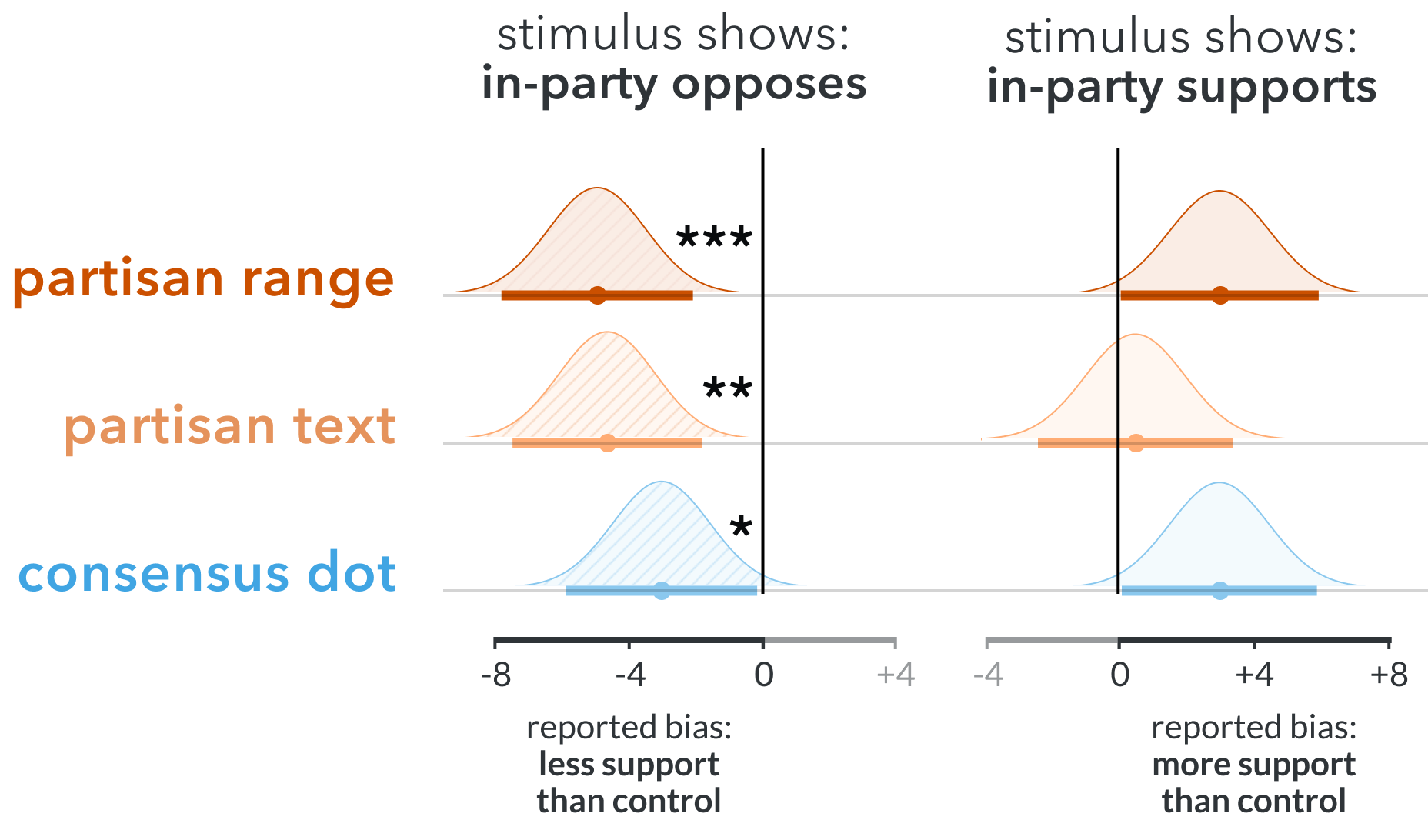}
 \caption{ Experiment 2 results: All charts showed aggregate social-conformity effects.
 When participants saw in-group opposition, they reported lower than expected support.  These plots show participants' reported attitude bias (x-axis: mean difference between reported attitudes for treatment minus \control{}, using estimated marginal means) after viewing charts showing their in-party opposing (left column) or their in-party supporting a given policy (right column). 
 Positive values indicate higher than expected support, negative indicate lower than expected. The uncertainty ranges indicate 95\% CIs. Stars indicate significant differences (* p<0.05\%, ** p<0.01\%, *** p=0.001\%). 
  }
 \label{pol3_exp2_intra_party_converge}
\end{figure}

To examine aggregate conformity toward a political party, we tested \partisandotrange{} and \partisantext{} against \control{}.  
We found the expected significant interaction between the (categorized) in-group value and the chart condition ($\chi^2$(2) = 17.2, p = 0.0002), suggesting collective conformity toward participants' in-parties (per \textbf{H2}). 
Dunnett-adjusted post-hoc comparisons (Fig.~\ref{pol3_exp2_intra_party_converge}) reveal this is driven by significant differences between the \control{} and \partisandotrange{} (MD = -4.9, p = 0.001, d = -0.19) and the \control{} and \partisantext{} (MD = -4.7, p = 0.002, d = -0.18), when the generated charts showed in-group opposition.
We also ran post-hoc comparisons between \partisandotrange{} and \partisantext{} and found no significant difference. The supplement reports the full statistics.

To examine aggregate conformity toward a national-identity group (US adults), we tested \consensusdotplot{} against \control{}. 
We found the expected significant interaction between the (categorized) in-group value and the chart condition ($\chi^2$(1) = 10.3, p = 0.001), suggesting collective conformity toward a national consensus (per \textbf{H2}). 
Dunnett-adjusted post-hoc comparisons (Fig.~\ref{pol3_exp2_intra_party_converge}) reveal this is driven by significant differences between \control{} and  \consensusdotplot{} (MD = -3.2, p = 0.03, d = -0.13) when the generated charts showed in-group opposition.
We also note that an alternative analysis of \consensusdotplot{}, modeling the in-group stimulus value as the hidden value generated for the participant's self-identified political party, might be a more analogous comparison between \consensusdotplot{} and the two partisan charts. We offer this analysis in the supplement and note that, under these assumptions, the interaction between the categorized stimulus value and the stimulus chart condition is not significant. The supplement reports the full statistics for both approaches.

\subsubsection{Social Conformity Discussion}
These results show that viewers' attitudes conform toward the visualized attitudes of their in-groups and that data visualizations can channel social-normative influences (supporting \textbf{H2}). 
As Fig.~\ref{pol3_exp2_stim_value_vs_reported_value_mean_diff} shows, for participants who saw \consensusdotplot{}, their responses were biased toward ``All US Adults.'' 
For those who saw \partisandotrange{}, their responses were biased toward their political in-party. 
Participants who saw \partisantext{} also biased toward their in-party, but this effect tapers or reverses at the extremes, possibly reflecting that \partisantext{} conveys categorical values (either ``favors'' or ``opposes''), whereas the visualizations show numeric values.

Both Fig.~\ref{pol3_exp2_stim_value_vs_reported_value_mean_diff} and Fig.~\ref{pol3_exp2_intra_party_converge} suggest a stronger effect for charts showing in-party opposition compared to in-party support. 
This is expected given the experiment design. Two of the five policies we tested are popular with both parties, but we showed them as much less popular than reality. 
For example, participants in the control condition reported strong bipartisan support (82-average) for requiring autonomous vehicles to have a human driver behind the wheel. However, in the treatment conditions, we showed this policy was opposed by Democrats (33-average stimulus values), supported by Republicans (66-average), and mixed nationally (50-average). Since the visualized reference points were typically lower than participants' prior attitudes, we would expect a social conformity effect to lean negatively for this policy.
 
The response curves in Fig.~\ref{pol3_exp2_stim_value_vs_reported_value_mean_diff} are consistent a discrepancy mechanic for attitude change \cite{sherif_social_1961,griffin_social_2012,bochner_communicator_1966}, where people change their attitudes in proportion to the discrepancy between their prior attitudes and an acceptable reference attitude. 
The x-axis in these charts can be interpreted as the discrepancy between participants' prior attitudes and the visualized reference attitude.
Given a moderately supportive (62-average) overall response from \control{} participants (a baseline estimate of overall prior attitudes), \consensusdotplot{} and \partisandotrange{} show that participants' attitudes changed in proportion to the stimulus value's distance from this baseline prior attitude.
% \omg{Through this lens, the tapering effects for \partisantext{} suggest that the "favor" or "oppose" categories have implied numeric values corresponding with the curves minima and maxima.}
The conformity effect for \consensusdotplot{} hints at a viable intervention for reducing policy polarization, since by definition these charts would pull people toward a more moderate national consensus.

\subsection{Experiment 2: Polarization and Framing}

% While H2 supports H3, H2 is insufficient to show collective polarization. T

We predicted that social conformity effects can build to group-level polarization (\textbf{H3}), which we define as between-party attitude divergence. We also predicted that these effects would depend on how the results are visualized (\textbf{H4}). 
Specifically, we predicted that partisan charts would polarize attitudes and common-ground charts would depolarize. 
While our previous analysis suggests potential effects of conformity, this may not necessarily imply polarization, which is a relative change in attitude distributions between two groups.

\begin{figure}[h!]
\centering
 \includegraphics[width = \columnwidth]{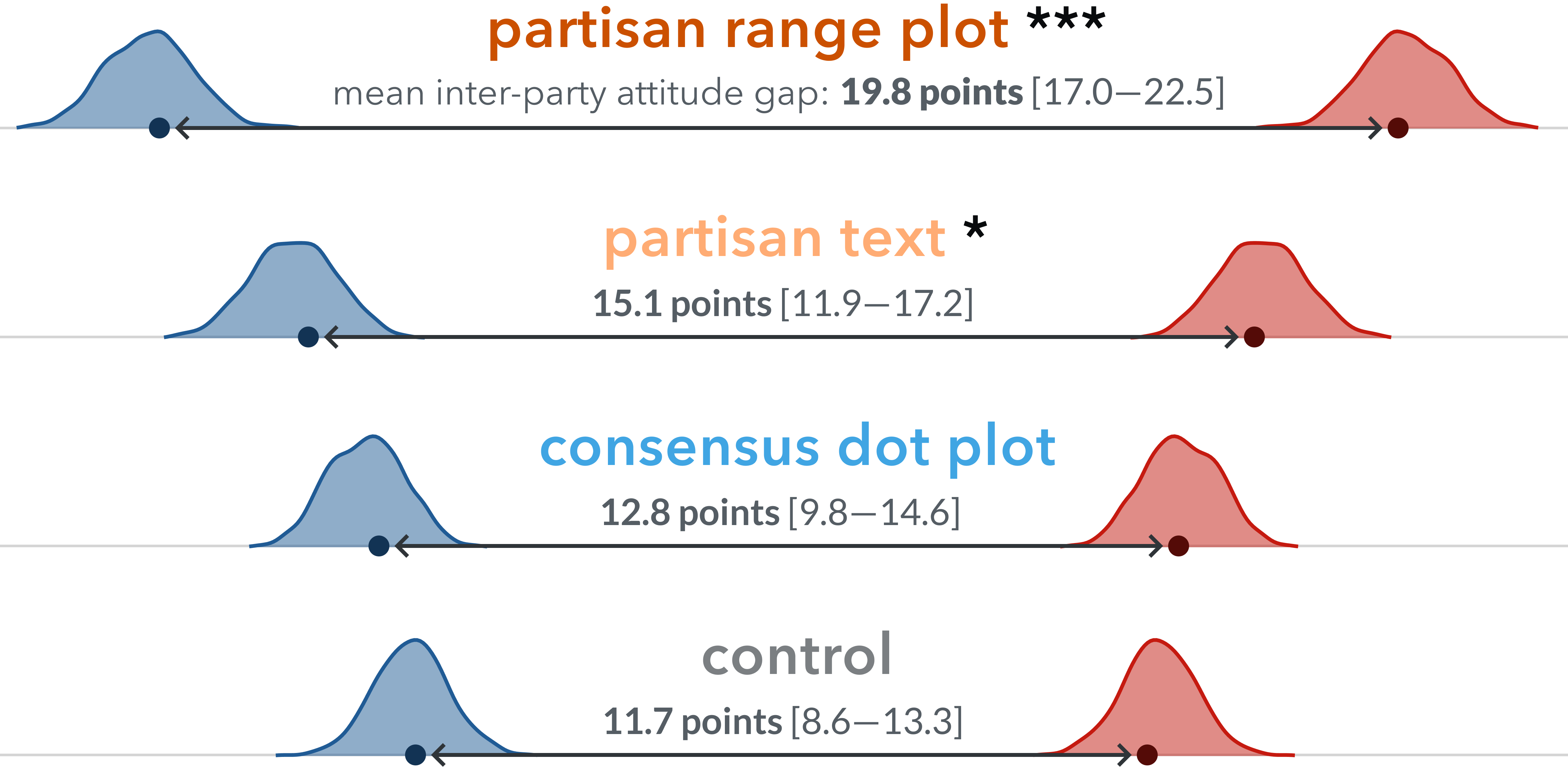}
 \caption{
 Experiment 2: \partisandotrange{} led to significantly more divergent polarization than the other three conditions. Horizontal bars show the mean inter-party attitude distance (gap) between left- and right-leaning participants. The symmetric distributions on the ends show the bootstrapped samples of how wide the bars could be. Bars are centered horizontally to avoid implying changes in absolute attitude positions for one particular party. Stars indicate significant differences-in-gaps from \control{} based on non-overlapping CIs (* = 95\%, *** = 99.9\%).}
 \label{pol3_exp2_cond_partisan_differences}
\end{figure}

Our dependent variable was the inter-party attitude gap, which is the average absolute difference in attitudes between left-leaning and right-leaning participants across all policies. 
We evaluate polarization for each treatment as the difference-in-gaps between the treatment and the \control{}. A treatment is polarizing if it shows a wider gap than \control{}, and depolarizing if the gap is smaller. 
We also evaluate the visualizations' unique impact on polarization (relative to categorical text descriptions used in prior political science work \cite{malka_more_2010, clifford_increasing_2021}) by comparing \partisandotrange{} to \partisantext{}.
While ``polarization'' has other colloquial or formal definitions, we focus on the inter-party attitude gap because it is more consistent with our experiment design. 
To clarify, this measure does not necessarily speak to radicalization or extremism (e.g.\ if two parties are both extreme and aligned, and an intervention moves one of them to more moderate attitudes, we consider this polarization as the two parties move away from each other).
The measure also does not speak to DiMaggio's "consolidation" (intra-party kurtosis) definitions of polarization \cite{dimaggio_polarization_1996}), which we would not expect to see given the randomly generated stimuli.

We analyzed the inter-party attitude gaps and the difference-in-gaps via bootstrapping.
We bootstrapped inter-party gaps for each condition by sampling the response data from the four conditions 10,000 times, with replacement, taking as many samples as responses. We calculated the inter-party gap as the absolute distance between each party's mean attitude, for each topic, then averaged the gaps across topics. We bootstrapped differences-in-gaps between the treatment and \control{} similarly by calculating inter-party gaps for the treatment condition and  \control{}, then subtracting the treatment's gap by the \control{}'s. We repeated this procedure for \partisandotrange{} vs \partisantext{}.

Fig. \ref{pol3_exp2_cond_partisan_differences} summarizes the inter-party gaps for each stimuli condition. We found that both \partisandotrange{} and \partisantext{} led to significantly wider gaps (more divergent polarization) than \control{} (supporting \textbf{H3}). The inter-party gap for \partisandotrange{} was 19.8 points (95\% CI = [17.0,22.5]), or 8.1 points wider than \control{} (MD = 8.1 95\% CI = [5.1, 12.3]). The inter-party gap for \partisantext{} was 15.1 points (95\% CI = [11.9,17.2]), or 3.4 points wider than \control{} (MD = 3.4, 95\% CI = [0.0, 7.1]). Differences between \consensusdotplot{} and \control{} were not significant. We also found that \partisandotrange{} led to significantly more polarization than \partisantext{}, with a 4.7 point wider gap (MD = 4.7, 95\% CI = [1.5, 9.0]).

\subsection{Experiment 2: Discussion}

The results suggest that public opinion visualizations exert a social-normative influence on viewers' attitudes (supporting \textbf{H2}), leading to social conformity effects for both national and partisan identity framings. 
This conformity effect can build to inter-party polarization (supporting \textbf{H3}) when visualized with partisan-framed charts like \partisandotrange{}. Given that both partisan treatments increased polarization, while \consensusdotplot{} did not, the results also imply framing effects for design choices (supporting \textbf{H4}). 
The wider inter-party gap for \partisandotrange{} vs \partisantext{} (19.8 vs 15.1 points) may suggest that visualizing polarization may be uniquely polarizing, at least relative to reporting similar gist information in text. However, this needs further research to differentiate between the effects of modality (text vs visualization) and specificity (``favor''/``oppose'' vs numeric values).
Even though \consensusdotplot{} led to significant social-conformity effects, 
and by definition it shows a moderate reference point, it did not reduce between-party polarization.

\begin{figure}[h!]
\centering
 \includegraphics[width = \columnwidth]{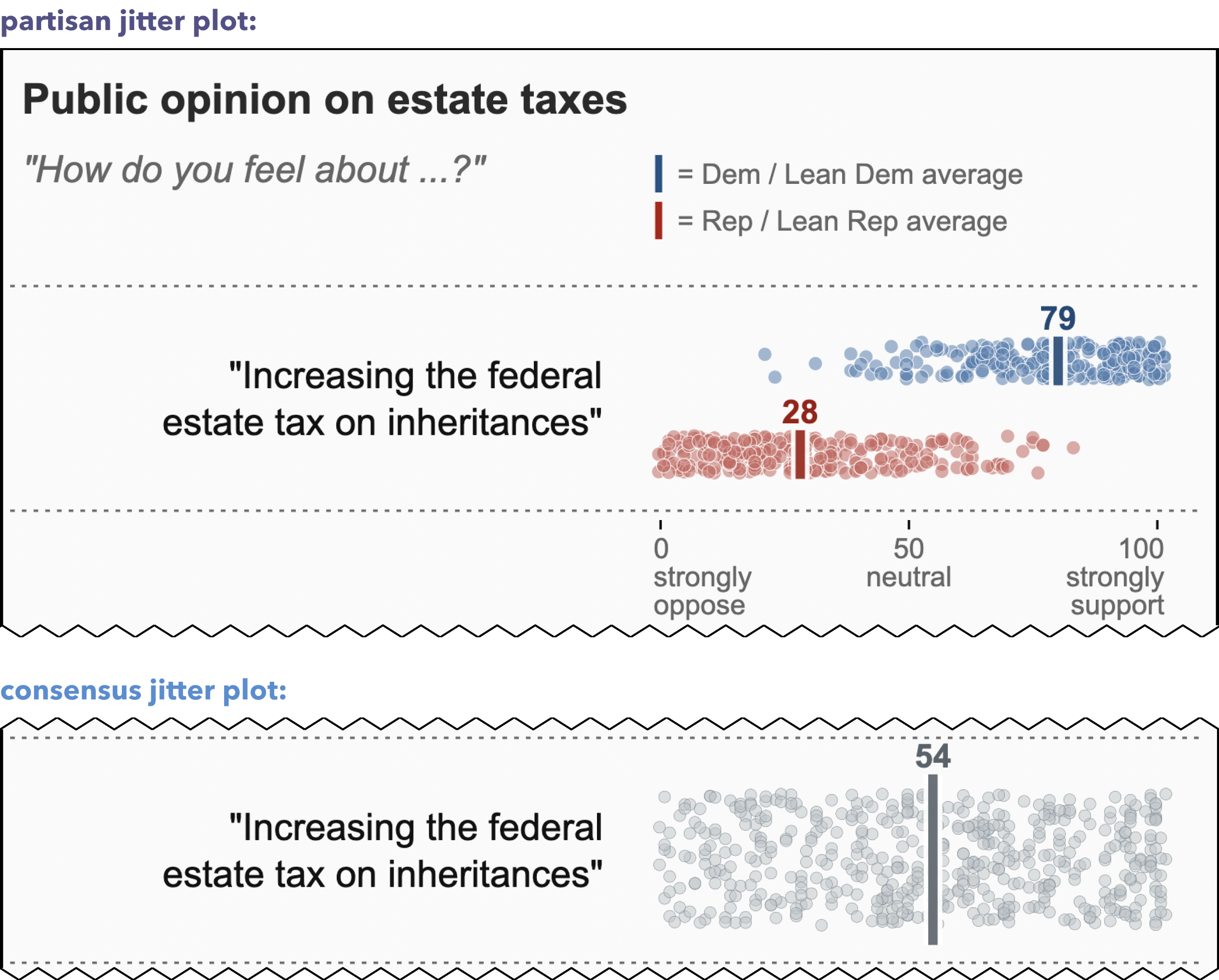}
 \caption{Experiment 3 conditions. Complete stimuli available on \href{https://osf.io/xqdw6/?view_only=8e614908f5ae42f794c650f753232c09}{OSF}.}
 \label{exp3_conditions}
\end{figure}

%%%%%%%%%%%%%%%%%%%%%%%%%%%%%%%%%%%%%%%%%%%
\section{Experiment 3}

While we have shown that partisan framing can polarize, we have had limited success with designs intended to depolarize. Experiment 1 showed that \consensusdotplot{} increased moderates' support for consensus gun-control policies. Experiment 2 showed that \consensusdotplot{} also exerts a social-conformity force, but had no effect on divergent polarization. In this final experiment, we look to other theories for design ideas that might depolarize policy attitudes (per \textbf{H4}).
In particular, we suspected that the dichotomization of partisan opinions may contribute to further polarization. 
Viewing out-parties as monoliths can increase harmful stereotypes, potentially influencing policy attitudes through affective polarization \cite{holder_dispersion_2023,druckman_affective_2021,ahler_parties_2018,lees_inaccurate_2020}. 
On the other hand, increasing perceived similarity \cite{balietti_reducing_2021} and correcting misconceptions of out-party composition and animus \cite{ahler_parties_2018, lees_inaccurate_2020} seem like viable interventions for depolarization. 
Through some combination of these mechanics, we suspected that highlighting the variance of intra-party attitudes may help depolarize viewers' attitudes.

\subsection{Experiment 3: Participants, Design, and Procedure}

We followed similar recruiting, exclusion protocols, experimental design, and procedures as the previous experiment, with a few changes. 
Instead of using MTurk's ``Premium Qualifications'' for partial partisan balance, we recruited until all cells had at least 125 participants, then randomly sampled 125 participants from each cell, for a fully-balanced design. 
This resulted in 750 participants total, 250 per condition, and 375 each of left- and right-leaning participants.
The slightly lower participant count reduced our expected power from 90\% to 87\%.
% Fewer participants than previous experiments (250 per condition, vs 274 previously) implies a slight loss of power (87\% vs 90\%), which we accepted given the unexpectedly slow rate of recruiting conservatives. }

\pheading{Stimuli Design:} We also tested two new designs (Fig. \ref{exp3_conditions}). \consensusjitterplot{} used a common-ground framing (showing ``U.S. Adults'') and jitter dots representing the individual attitudes of several hundred (simulated) U.S. Adults. We suspected that showing a breadth of attitudes may additionally dispel narratives Americans' attitudes are irreconcilably divided. 
\partisanjitterplot{} used a partisan-group framing, showing average support for either party, combined with jitter dots representing the individual attitudes of several hundred (simulated) partisans. 
We suspected that, if illusory dichotomization were an issue, explicitly showing the overlap between parties may help dispel exaggerated polarization beliefs and minimize inter-party stereotypes. 
We generated group-level stimulus values for the parties and U.S. adults following the procedure from Experiment 2. We then sampled an additional 250 values per group, from matching beta distributions, and used these values to place the jitter dots. 
To create a more fair test for depolarization, we also used only the two most partisan policy topics from Experiment 2: increasing estate taxes (liberal-favored) and expanding nuclear energy (conservative-favored).

\begin{figure}[h!]
\centering
 \includegraphics[width = \columnwidth]{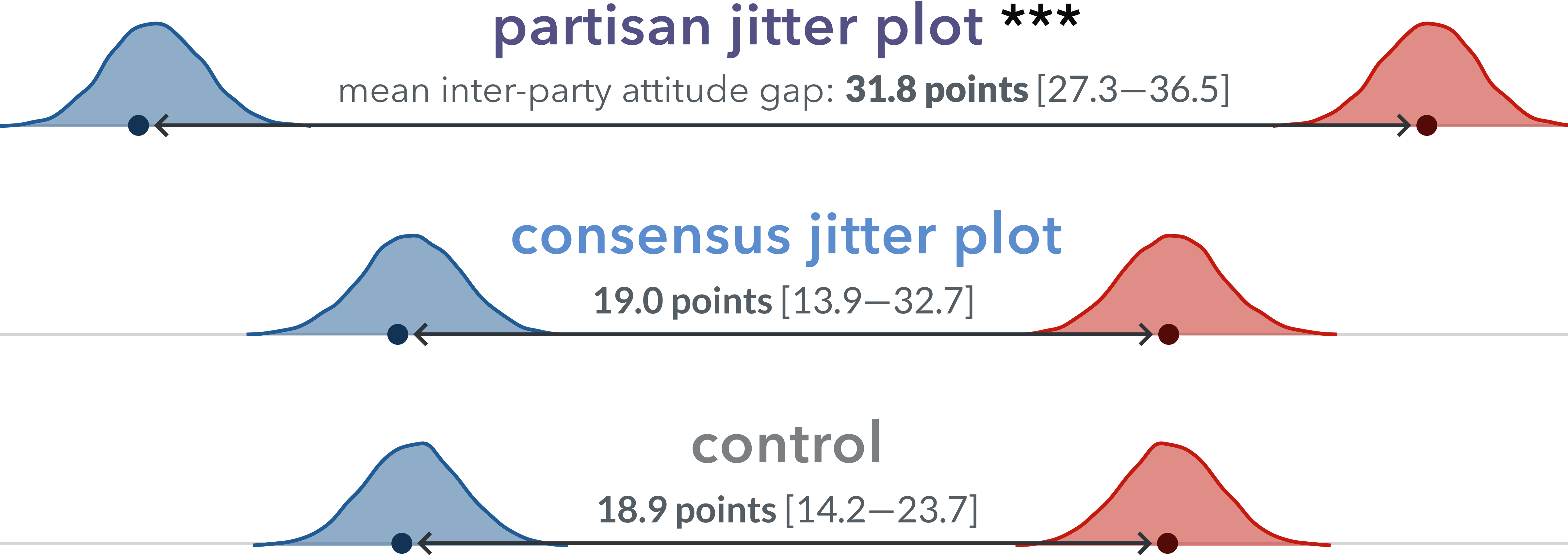}
 \caption{
 Experiment 3: Neither design reduced polarization. \partisanjitterplot{} made it worse. See Fig. \ref{pol3_exp2_cond_partisan_differences} for chart interpretation notes.
  }
 \label{pol3_exp3_cond_partisan_differences}
\end{figure}

\subsection{Experiment 3: Results}

We predicted that charts showing within-group variance could depolarize participants' policy attitudes (per \textbf{H4}).
Following our analysis from Experiment 2 to evaluate between-party attitude divergence, Fig.~\ref{pol3_exp3_cond_partisan_differences} shows results from bootstrapping inter-party attitude gaps. 
We found that the inter-party gap for \partisanjitterplot{} was 31.8 points (95\% CI [27.3,36.5]), or 12.9 points wider than \control{} (MD = 12.9, 95\% CI [6.2, 19.7]). Differences-in-gaps between \consensusjitterplot{} and \control{} were not significant. 
% \omg{We repeat the complete Experiment 2 analysis in the supplement.}

\subsection{Experiment 3: Discussion}

While visually dichotomizing groups contributes to harmful stereotypes, and out-party stereotyping contributes to polarization \cite{holder_dispersion_2023, ahler_parties_2018, lees_inaccurate_2020}, our attempts to highlight intra-party variation in political attitudes had unexpectedly polarizing effects. Instead of improving polarization, highlighting intra-group attitude variance either did not help (\consensusjitterplot{}) or made it much worse (\partisanjitterplot{}), corroborating findings from Experiment 2.

For \partisanjitterplot{}, there could have been a number of additional factors to consider, such as the relationship with affective polarization as a mediator, sensitivity to the range of values (i.e.\ showing individual dots implies showing people on both extremes, which may imply even more extreme social norms), or perhaps just that it is even more red and blue ink on the page to activate partisan identities. \consensusjitterplot{} results were similar to \consensusdotplot{}: The treatment did not help polarization, but it did not hurt. 

While \partisanjitterplot{} (Fig.~\ref{pol3_exp3_cond_partisan_differences}) shows a wider gap compared to \partisandotrange{} from Experiment 2 (Fig.~\ref{pol3_exp2_cond_partisan_differences}), this reflects differences between experiments, not differences between the charts. Experiment 2 included responses to three additional policies with more baseline bipartisan support, dampening polarization for all conditions.

\section{Experiment Summary}
Our experiments show a path from social conformity to polarization.

\pheading{In Experiment 1}, we show that visualizing policy opinions can shape policy opinions, particularly for moderate partisans. 
We used a realistic visualization of gun policy attitudes to show that a popular chart type is not merely a passive source of political information; rather, it can actively shape our politics.

\pheading{In Experiment 2}, we showed that data visualization can induce social conformity, and as a consequence, visualizing polarization can increase polarization.
We also found that visualization can have a stronger influence on partisan attitudes than the verbal gist summaries studied in political science.

\pheading{In Experiment 3}, our designs, intended to decrease polarization, either made it worse or had no effect, replicating our previous results.

%%%%%%%%%%%%%%%%%%%%%
\section{Discussion and Design Implications}
Our results establish two effects between social psychology and data visualization: 1) social conformity in attitudes can be induced through data visualization and 2) visualizing political polarization can accelerate it. 
The effects further question visualizations' presumed neutrality. Other work questions rhetorical neutrality \cite{lee-robbins_affective_2023, hullman_rhetoric_2011} or suggests that causally-neutral visualizations lead to unsupported beliefs \cite{d2020data, holder_dispersion_2023}. These results question neutrality-as-passivity, showing that visualizations can actively shape the phenomenon that they portray.

In recent debates on newsroom neutrality, one dominating view encourages journalists to use intentionally narrow criteria when deciding what to publish (``Is it true?\ Is it important?'') and be ``profoundly skeptical'' of subjective concerns about harmful outcomes \cite{sulzberger2023}. 
These cultural norms can affect data journalists, who are aware of the persuasive influence of their work, but fear stigmatization if they ``talk openly'' about it \cite{lee-robbins_affective_2023}. 
Our results highlight the potentially harmful consequences for publishing partisan polling results, a popular news topic. 
They corroborate others' suggestions that stories of polarization may be self-perpetuating \cite{cohen_party_2003,  levendusky_media_2016, klein_why_2020}. Together, this suggests not only a need for more open discussion but also provides grounds for reconsidering a default posture of consequential skepticism.

We extend prior political science work \cite{clifford_increasing_2021, malka_more_2010}, showing that partisan cue effects translate to data visualization. 
We also show that visualization can have a stronger effect than the verbal, categorical descriptions of partisan support used in these prior experiments, suggesting that visualization (or, at least, numerically represented attitudes) can add a unique dimension to established partisan-conformity effects.

Most urgently, the conformity effects may have consequences for public health communication. To promote health equity, institutions often publish charts highlighting adherence disparities between social groups (e.g.\ differences in vaccination rates). To the extent that viewers interpret these disparities as representing groups' normative attitudes (e.g.\ ``vaccines are unpopular with people like me''), these charts may backfire and further reduce support within the (marginalized) communities they are intended to help. Further research on social conformity in public health visualizations is an urgent next step.

\pheading{Design Implications:} Designers visualizing social norms should understand that they influence their audiences, not just inform them. 
Choosing to show that an attitude is popular (or not) within a certain social group can make it more (or less) popular for that group. 
This implies that, when deciding what to visualize, whatever benefits readers might gain from improved sense-making should be offset by the risks (or benefits) of influencing their attitudes (e.g.\ further polarization).

%%%%%%%%%%%%%%%%%%%%%%%%%%%%%%%%%%%%%%%%%%%
\section{Limitations and Future Directions}

As an initial investigation of polarization through visualization, we discuss several limitations and suggest directions for future research.

\pheading{Other visualizations.} While we found that popular visualization approaches can increase polarization, we have yet to find designs that can reduce it. 
We suspect that other non-partisan aggregations (e.g.\ by geography, perceived expertise, other social affiliations), the salience of partisan branding (e.g.\ blue donkeys for Democrats), the visual range of markers across the scale, or other ways of conveying common ground between parties may still be viable interventions. 

\pheading{Other polarization dynamics.} We focused on polarization as inter-party divergence, but other definitions may reveal other dynamics. We suspect Dimaggio's consolidation definitions might clarify a path toward depolarizing visualizations \cite{dimaggio_polarization_1996}.

\pheading{Other topics.} Experiments 2 and 3 used relatively non-partisan topics as case studies, referencing examples used in prior work \cite{pew_driverless_2017,pew_nuclear_2022,clifford_increasing_2021}.
However, we found that even these showed baseline polarization. 
Exploring truly neutral topics may help clarify how opinion polling visualizations contribute to the early stages of a topic becoming polarized (e.g.\ masking, prior to Covid-19). There may also be asymmetries found in different framings of the same topic, for example where participants may support "increasing gun restrictions" more than they oppose "increasing gun access."

\pheading{Individual differences.} While the experiment was not intended to explore individual differences, our results showed that demographic factors could be significant (see the supplementary materials). 
Future work can explore the impacts of prior policy-related beliefs, affective polarization, actual political knowledge (vs self-reported familiarity), or needs for security, certainty, and individuation \cite{federico_when_2021}.

\pheading{Positive vs negative partisanship.} Negative partisanship (i.e.\ biasing away from an out-party) is often cited as more influential than positive partisanship (i.e.\ biasing toward an in-party). 
So while we might expect participants to be more repulsed by their out-party than attracted to their in-party, our experiments were not designed to differentiate between these two forces. Future work could explore even more obscure topics to isolate these effects.

\pheading{Post-only design.} Our experiments only gathered participants' attitudes and political alignment after they viewed the stimuli. To understand individual effects (e.g.\ how many people change their attitudes and by how much?), future studies could use repeated measures designs or explore other interesting attitude proxies, such as collecting predicted attitudes for \textit{other} people. 

\pheading{United States focus.} While political polarization exists around the world, this study was designed specifically for participants encultured within the United States, including the topics chosen and the political parties featured in the visualizations. 
Further work is needed to see if these results generalize to other countries and cultures.
The U.S. is also a predominantly two-party system; future work could also explore implications for multi-party politics.

% leave out until camera ready
\acknowledgments{
We are grateful to the participants who made this study possible and our reviewers for their insightful feedback. 
We also thank Jeremy Wilmer and Sarah Kerns for their advice and encouragement, as well as Amanda Holder for her support and infinite patience.
This work was partially supported by NSF award IIS-2237585.}
 % This work was supported by NSF award CISE HCC-223758

\clearpage
%% BIBLIOGRAPHY %%
% \balance
\bibliographystyle{abbrv}
\bibliography{reference}

\end{document}